\documentclass[aps,prl,twocolumn,groupedaddress]{revtex4}
\usepackage{graphicx}
\usepackage{amsmath,amsfonts,amssymb}
\usepackage{color}
\usepackage{float}
\begin{document}

\title{From Planck area to graph theory:\\
Topologically distinct black hole microstates}

\author{Aharon Davidson}
\email{davidson@bgu.ac.il}
\homepage{https://physics.bgu.ac.il/~davidson}
\affiliation{Physics Department, Ben-Gurion University
of the Negev, Beer-Sheva 84105, Israel}

\date{july 7, 2019}

\begin{abstract}
	We postulate a Planck scale horizon unit area, with no
	bits of information locally attached to it, connected but
	otherwise of free form, and let $n$ such geometric
	units compactly tile the black hole horizon. 
	Associated with each topologically distinct tiling
	configuration is then a simple, connected, undirected,
	unlabeled, planar, chordal graph. 
	The asymptotic enumeration of the corresponding
	integer sequence gives rise to the Bekenstein-Hawking
	area entropy formula, automatically accompanied
	by a proper logarithmic term, and fixes the size of the
	horizon unit area, thereby constituting a global realization
	of Wheeler's "it from bit" phrase.
	Invoking Polya's theorem, an exact number theoretical
	entropy spectrum is offered for the 2+1 dimensional
	quantum black hole. 
\end{abstract}

%\pacs{}

\maketitle
\section{I. Introduction}

The semiclassical Bekenstein-Hawking black
hole area entropy formula \cite{BH}
\begin{equation}
	S_{BH}=k_B\frac{A_{BH}}{4\ell_P^2} ~,
	\label{SBH}
\end{equation}
governed by the horizon surface area $A_{BH}$,
measured in Planck units $\ell_P^2=G\hbar/c^3$, and
factorized by the Boltzmann constant $k_B$, is still 
as mysterious as ever.
We have no compelling answer for what the physical
degrees of freedom underlying the
Schwarzschild black hole prototype actually are, or for how
to identify and count its elusive quantum
microstates.
Various attempts to address this issue
have come from all corners of theoretical physics, way
beyond general relativity,
including string theory \cite{string}, loop
quantum gravity \cite{loop}, and AdS/CFT \cite{AdSCFT},
each theory contributing its inimitable insight.
According to Maldacena \cite{Malda}, 
"these microstates do not have an explicit calculable
description within the regime that gravity is a good
approximation."

It was Bekenstein \cite{nA} who first realized that the
black hole surface area may serve as a classical adiabatic
invariant, and as such, it must exhibit a discrete ladder
spectrum of the form $A_{BH}(n)=n A_1$.
This has opened the door for a variety of
Bekenstein-Mukhanov \cite{BM} inspired quantum black
hole models \cite{models},
the majority of which assume $\gamma$ (a natural
number) bits of information locally encoded in each
Planck area on the horizon.
Such a local realization of Wheeler's 'it from bit' phrase
\cite{Wh} gives rise to a total of $g(n)=\gamma^n$
configurations.
However, no compelling clue was given as to what these
bits actually stand for or what physics is capable of
hosting them on the event horizon.
Along these lines, it is worth recalling the 't Hooft-Susskind
holographic principle \cite{holo} which asserts that all of
the information contained in some closed region of space,
saturated by Eq.(\ref{SBH}), can in fact be represented as
a hologram on the boundary of that region.

While the general idea of a fundamental Planck scale
horizon unit
area is not new, the role it plays in the present model
is novel.
In fact, in contrast to almost all Bekenstein-Mukhanov-
type models, no bits of information are locally attached
to any single unit area.
An individual Planck area does not play any local role
at all here.
Alternatively, our interest is focused on a collective
mode of all Planck units involved, with the various
topologically distinguished configurations highly
resembling (and perhaps identified as) the quantum
black hole microstates.
Their counting, and the subsequent recovery of
Eq.(\ref{SBH}) in the semiclassical limit, which is automatically
accompanied by a proper logarithmic term, is carried
out by invoking graph theoretical enumeration.
Triggered by graph theory, the black hole discrete
entropy spectrum is furthermore shown to establish a
serendipitous link with number theory (with the focus
on Polya's theorem \cite{Polya}).

\section{II. Horizon tiling}

The main ingredient in our quantum black hole model
is a postulated Planck size horizon unit area
\begin{equation}
	A_{P}=\eta \ell_P^2 
	\label{Planck}
\end{equation}
where $\eta$ is a dimensionless universal constant,
which is eventually fixed by means of graph theory.
Eq.(\ref{Planck}) may further serve as a geometric
lower bound inspired by the 't Hooft-Susskind
holographic principle \cite{holo}, but this is beyond
the scope of the present model.
Based on self-consistency grounds, the Planck unit
area must exhibit a locally connected structure but
can otherwise take any free form.
Its boundary can thus undergo any arbitrary variation
as long as the size of the surrounding area is preserved
in accordance with Eq.(\ref{Planck}).

We now attempt to compactly tile the
black hole horizon surface area $A_{BH}$ by exactly
\begin{equation}
	n=\frac{A_{BH}}{A_P}
	\label{n}
\end{equation}
elementary Planck unit areas.
It makes no sense, and actually there is no option, to perform
this uniformly.
The reason is quite obvious:  While Planck unit areas are
all topologically equivalent, they may still differ from each
other by acquiring arbitrary, albeit connected, shapes.
For any given integer number $n$ of Planck unit areas,
the relevant question is then how many topologically
distinct tiling configurations $g(n)$ actually exist?

Counting configurations calls for graph theory enumeration.
The first step is to show, by construction, that associated
with each topologically distinct tiling configuration
is a certain mathematical graph, defined as a set of vertices
connected by edges.
There are four simple instructions:
\begin{enumerate}
\item Assign a graph vertex to each Planck unit area, and locate
this vertex at some point on that unit (this is always possible
due to the local connectedness).
\item Connect any two such vertices by a graph edge if and only
if the two corresponding Planck unit areas touch each other.
\item Draw the graph on the horizon itself, and take into
account the fact that from the topological point of view,
as a 2-dimensional spherical surface $S^2$ with no handles,
the Schwarzschild horizon is of genus $0$.
This is a guaranteed by Hawking theorem \cite{Hawking}
which holds for asymptotically flat 4-dimensional black holes
obeying the dominant energy condition.
Genus dependence will be briefly discussed later.
\item By choosing one graph face and puncturing a hole in it,
one may further, via a stereographic projection, reliably
transform the graph from the sphere onto a plane.
The punctured face on the sphere becomes the exterior
face on the plane. 
\end{enumerate}

\begin{figure}[h]
	\center
	\includegraphics[scale=0.4]{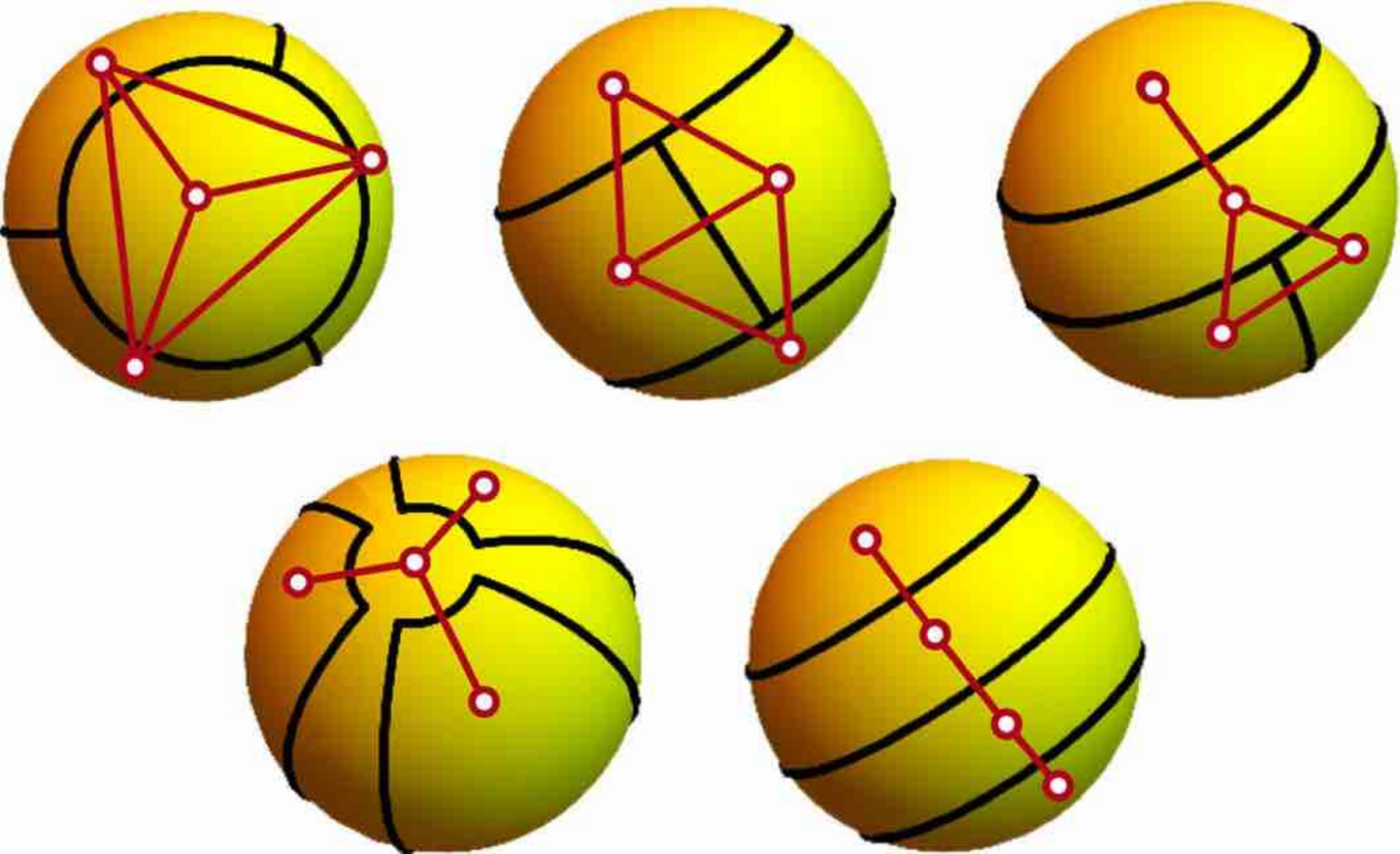}
	\caption{Translating horizon tiling into graph
	theory language.
	The demonstration is carried out for $n=4$
	deformable Planck unit areas (separated by black
	borders), resulting in $g(4)=5$ topologically
	distinct configurations.
	Associated with each such configuration
	is a simple, connected, undirected, unlabeled,
	planar, chordal graph (plotted in red).}
	\label{Tiling}
\end{figure}

The transition from the black hole horizon tiling to
graph theory is demonstrated in Fig.\ref{Tiling} for
$n=4$ vertices.
%% Fig Tiling %%%
%%%%%%

\section{III. Graph theory}

\begin{figure}[h]
	\center
	\includegraphics[scale=0.51]{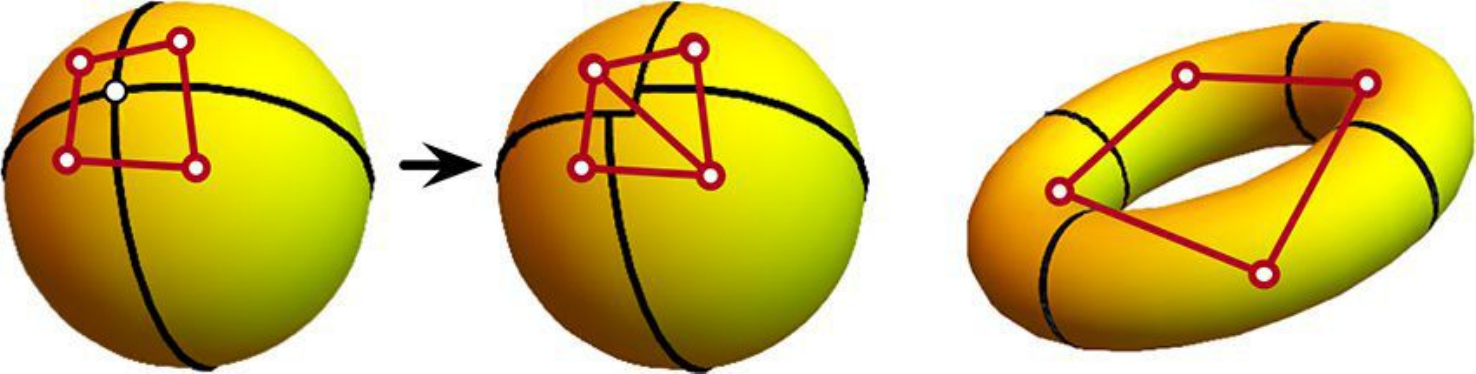}
	\caption{The 4-edge square graph, representing
	a truncated (cutoff poles) beach ball, is excluded. 
	The 4-meeting point on the ball is unstable against
	small shape variations of the horizon unit areas,
	bifurcating into two 3-meeting points.
	This is translated into graph theory as adding a
	chord.
	On a torus, as a counterexample, the square graph
	is permissible.}
	\label{Square}
\end{figure}

Prior to performing enumeration, we must accurately
specify what kind of graphs we are actually dealing with.
By construction, mostly on geometric or physical grounds,
these graphs must be as follows:
\begin{enumerate}
\item {\it{Simple}}. - The graph cannot contain
loops and/or multiple edges.
A Planck area unit does not touch itself, and it is clear
whether or not two Planck areas share a common border.
\item {\it{Connected}.} -  There must be a path
from any vertex to any other vertex of the graph.
Allowing for a disconnected graph, an isolated Planck
area unit for example, would mean leaving a region of
\item {\it{Undirected}}. - No flow is described in
the model.
In turn, no arrows need to be attached to the graph edges.
\item {\it{Unlabeled}}. - Reflecting the fact that
individual Planck areas have no distinct identifications
except through their interconnectivity, the graph vertices
do not carry any serial numbers.
As we shall see, this is the strongest requirement on our list.
On the practical side, it is much harder to enumerate
unlabeled than labeled graphs.
\item {\it{Planar}}. - A graph is planar if it can be
drawn in a plane, or on a handle-free sphere like the horizon,
without graph edges crossing.
Be aware that
(i) fake edge crossings can be removed by replacing straight
lines by Jordan arcs, and
(ii) there may be several representations of the same
planar graph.
For any given number $n$ of nodes, the number of labeled
planar graphs turns out to be much larger than the number
of unlabeled planar graphs since almost all planar graphs
have a large automorphism group.
\item {\it{Chordal}}. -  A chordal graph, also called
a  triangulated graph, is a simple graph in which every
cycle of more than three vertices has a chord ($=$ an edge
that is not part of the cycle but connects two vertices of the
cycle).
Beware that chordality is sometimes visually hidden.
To see why this is relevant for our case, let four Planck
areas meet at some point on the horizon.
However, such a configuration turns out,  to be topologically
unstable with respect to small variations in the shapes of
the Planck areas involved.
Roughly speaking, a 4-meeting point easily bifurcates into
two 3-meeting neighboring points, which is translated into
graph theory as adding a chord.
The corresponding disqualification of the square graph
is illustrated for $n=4$ in Fig.\ref{Square}.
To sharpen the genus dependence, note that when plotted
on a torus (genus $1$), rather than on a sphere (genus $0$),
the square graph becomes stable and thus permissible. 
\end{enumerate}

%% Fig Square %%%
%%%%%%

\begin{figure}[ht]
	\center
	\includegraphics[scale=0.35]{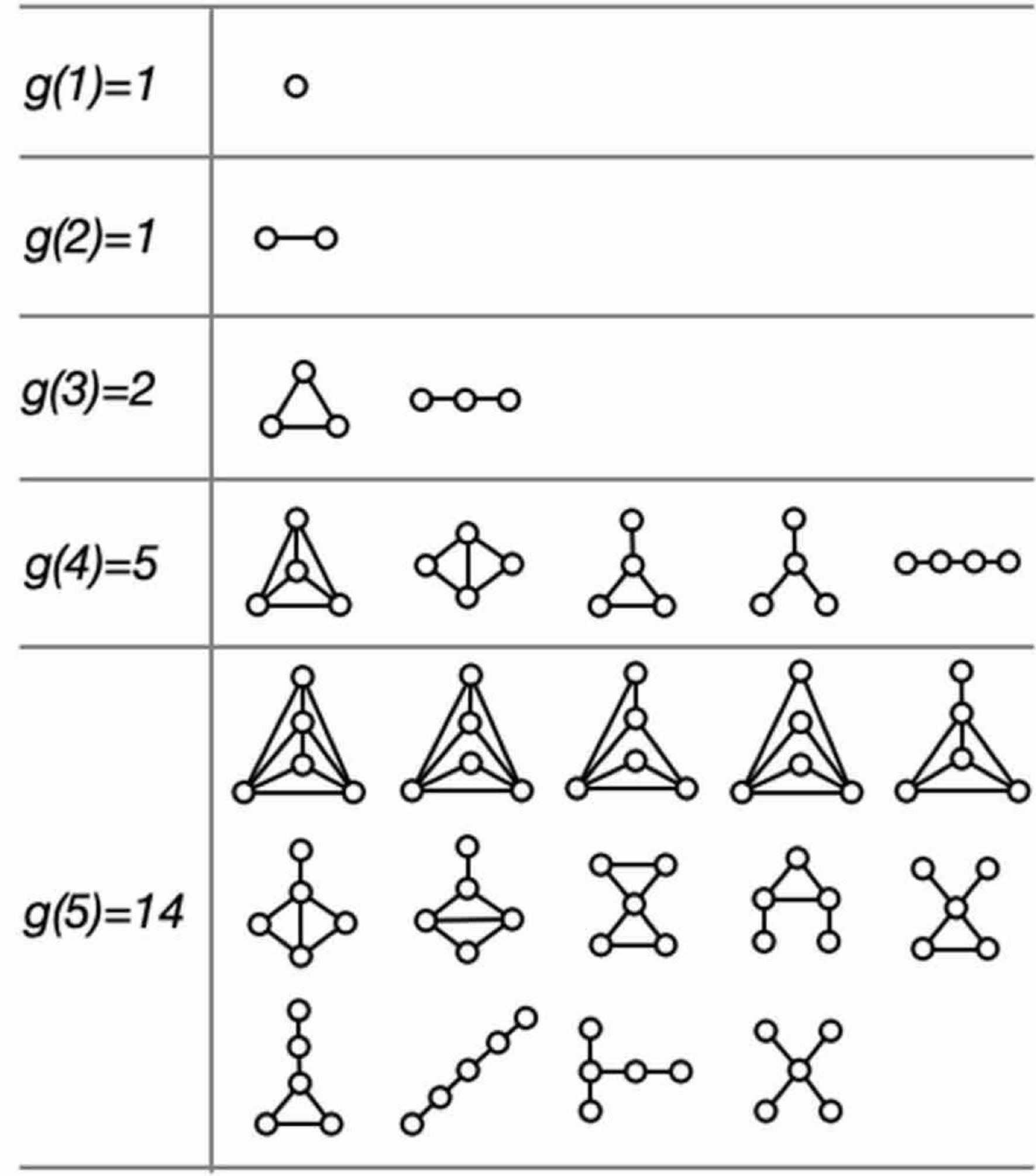}
	\caption{The integer sequence OEIS A243787:
	Simple, connected, undirected, unlabeled, planar,
	chordal graphs with $n$ nodes. 
	The inner structure of these graphs is solely
	composed of triangles and trees.
	In our model, each graph represents a topologically
	distinct black hole microstate.}
	\label{g(n)}
\end{figure}

Altogether, the above list of graphic features homes in
on a particular integer sequence classified as OEIS A243787.
To be more explicit, the first terms of the series (so far, only
the first 14 terms have been calculated \cite{Huffner})
are given by
\begin{equation}
	g(n) = 1, 1, 2, 5, 14, 52, 228, 1209, ...
	\label{gn}
\end{equation}
(See Fig.\ref{g(n)} for the graphs associated with the first terms.)
It starts like the Catalan series, but then grows faster.
For comparison, had we given up the chordality requirement,
we would end up with a much larger set
\begin{equation}
	g_{u}(n) = 1, 1, 2, 6, 20, 99, 646, 5974, ...
	\label{gun}
\end{equation}
of unlabeled connected planar graphs.
Clearly, the former Eq.(\ref{gn}) is a subsequence of the
latter Eq.(\ref{gun}).

%% Fig g(n) %%%
%%%%%%

Treating all topologically distinct configurations on equal 
footing, with each individual configuration serving as a
distinct quantum mechanical microstate, the statistical black
hole entropy is given by the Boltzmann formula
\begin{equation}
	S_{BH}=k_B \log g(n) ~.
	\label{Boltzmann}
\end{equation}
As anticipated, the lightest Schwarzschild black hole,
carrying mass $m_{1}=m_P \sqrt{\eta/16\pi}$, comes with
a vanishing entropy $S_{1}=0$.
The nontrivial microstate degeneracy starts at $n=3$.
An exact analytic formula for $g(n)$ is still unknown, but
some efficient enumeration algorithms do exist.
However, at this stage, this is not
what really matters.
Bearing in mind that the fate of our model primarily depends
on making contact with Eq.(\ref{SBH}) at the large-$n$
semiclassical limit, we content ourselves with an asymptotic
enumeration formula.

\section{IV. Asymptotic enumeration}

Counting labeled planar graphs appears to be much
easier than counting planar unlabeled graphs. 
The asymptotic number $g_{l}(n)$ of labeled planar graphs
has been shown, following a superadditivity argument
\cite{McDiarmid}, to obey the limit
\begin{equation}
	\lim_{n\to\infty} (g_{l}(n)/n! )^{1/n} 
	\rightarrow \gamma_{l}~.
	\label{liml}
\end{equation}
Upper as well as lower bounds on the constant $\gamma_{l}$
were numerically derived, but the final word was given 
analytically by Gimenez and Noy \cite{Noy}.
To be more specific, they calculated
\begin{equation}
	g_{l}(n) \simeq
	\alpha_{l}n^{-\frac{7}{2}}\gamma_{l}^n n! ~,
	\label{Noy}
\end{equation}
where $\alpha_{l}\simeq 0.43 10^{-5}$ and
$\gamma_{l}\simeq 27.23$.
As far as the unlabeled planar graphs are concerned,
owing to their large exponential number of automorphisms,
the limit on the corresponding asymptotic number
$g_{u}(n)$ of configurations is conceptually different.
In fact, it has been shown \cite{Denise} that
\begin{equation}
	\lim_{n\to\infty} g_{u}(n)^{1/n} 
	\rightarrow \gamma_{u}~,
	\label{limu}
\end{equation}
thereby consistently defining $\gamma_{u}$ as the unlabeled
planar graph growth constant.
Notice that, in comparison with Eq.(\ref{liml}),  the $n!$ factor
is gone.
In turn, with Eq.(\ref{Boltzmann}) in mind, the leading linear $n$ behavior of
$\log g_{u}(n)\simeq n\log \gamma_{u}$ is crucial for ouir model, to be contrasted
with the problematic (for our needs) leading behavior of
$\log g_{l}(n)\simeq n\log n$.
Apart from the $n!$ factor, the asymptotic enumeration of
unlabeled planar graphs cannot be too different analytically
from that of labeled planer graphs.
Thus it comes with no surprise that, in analogy with
Eq.(\ref{Noy}), Gimenez and Noy have derived
\begin{equation}
	g_{u}(n) \simeq
	\alpha_{u}n^{-\frac{7}{2}}\gamma_{u}^n ~,
	\label{xxx} 
\end{equation}
for some $\alpha_{u},\gamma_{u}$.
At this stage, while the exact value of $\gamma_u$ is still
unknown, Bonichon {\it{at al.}} \cite{Bonichon} have closed
the range to $27.23<\gamma_u<30.06$.

For the sake of enumeration, it is useful to probe the inner
structure of the graphs involved.
In our case, one starts from a subset of so-called
maximal planar graphs, which are nothing but triangulations.
For a given number $n$ of vertices, they exhibit $(3n-6)$
edges and $(2n-4)$ faces.
The corresponding integer sequence OEIS A000109 is
given by $g_{\triangle}(n)=0,0,1, 1, 1, 2, 5, 14, 50, ...$.
No new edges can be added without violating planarity.
All the other graph members in our list, for the same
given $n$, can now be manually constructed by removing
edges, one by one.
In doing so, however, one has to be careful
 (i) to maintain graph connectedness, and
 (ii) to create no holes, in the chordal sense explained earlier.
 The edge removal process divides the various $n$ graphs
 into $\{n,k\}$ subcategories for $k=0,1, ...,2n-5$, with
 $\sum_k g(n,k)=g(n)$.
For example, $g(5,k)=1,1,3,3,3,3$ for $k=0,1,...,5$,
respectively, with $\sum_k g(5,k)=14$.
 The number of edges and faces in the $\{n,k\}$ level is
$e=3n-6-k$ and $f=2n-4-k$, respectively.
Thus, the physically allowed graphs will have only
triangles and trees as their inner building blocks, an
important observation for enumeration purposes.

As anticipated, the asymptotic enumeration of our simple,
connected, undirected, unlabeled, planar, chordal graphs
is of the generic form
\begin{equation}
	g(n) \simeq
	\alpha n^{-\frac{5}{2}}\gamma^n ~.
	\label{enum}
\end{equation}
The exact value of the graph growth constant $\gamma$
has not been calculated yet. 
However, strict bounds on $\gamma$ do exist, an upper
bound as well as a lower bound (see below).
The factor $n^{-5/2}$ deserves special attention.
It is notably different from the analogous factor of
$n^{-7/2}$ [see Eqs. {\ref{Noy} and \ref{xxx}}] which
characterizes planar but not necessarily chordal graph
enumeration, to be regarded \cite{Bodirsky} as a direct
consequence of the triangle and tree composition of the
graphs involved.
For comparison, had we dealt with rooted tree graphs,
we would have obtained $n^{-3/2}$. 
Note in passing that graph enumeration is genus dependent.
Had the horizon been genus $g$, the counting function 
$g(n)$ would have been slightly modified \cite{genus},
\begin{equation}
	g(n) \simeq
	\alpha n^{\frac{5(g-1)}{2}}\gamma^n ~.
\end{equation}
The situation gets even trickier if the topology
includes an $S^1$ factor whose chirality (clockwise and
anticlockwise directions) opens the door for directed
graphs.

\section{V. Black hole entropy}

Altogether, the semiclassical large-$n$ asymptotic
expansion of the corresponding Boltzmann entropy
Eq.(\ref{Boltzmann}), is then given by
\begin{equation}
	S_{BH}(n)= k_B (n \log{\gamma}
	- \frac{5}{2}\log{n}+...)
	\label{log}
\end{equation}
Appreciating the linear-$n$ behavior of the leading term,
the connection with the Bekenstein-Hawking formula
Eq.(\ref{SBH}) can finally be established provided one
identifies
\begin{equation}
	\eta=4\log{\gamma}~.
\end{equation}
Note in passing that in our case, unlike in the
Bekenstein-Mukhanov model, there is \textit{a priori} no need
for $\gamma$ to be an integer.
It is by no means trivial that the exact size of the
horizon unit area, considered to be a purely (quantum
gravitational) geometrical feature, gets fixed by means
of graph theory.
In the present model, the latter conclusion is rooted in
the assumption that the fundamental horizon unit areas
are locally indistinguishable from each other, an assumption
which is translated into unlabeled rather than labeled
graphs.
This is a critical point.
Had we dealt with labeled graphs, we would have faced
the disastrous behavior $\log g_l(n)\simeq n\log n$
and never recover the Bekenstein-Hawking limit.

At this stage, the exact value of the graph growth
constant $\gamma$, crucial for fixing the Planck area
unit in Eq.(\ref{Planck}), is only known to lie in the range
\begin{equation}
	9.48<\gamma<30.06 ~~\Longrightarrow~~
	8.98<\eta<13.61 ~.
\end{equation}
It is an order of magnitude larger than the popular
values of $\gamma=2,3,4$ which we see in
Bekenstein-Mukhanov-inspired models.
The lower bound \cite{Tutte} reflects the fact that
our graphs contain all unlabeled triangulations as a subset.
Smaller subsets include the pure trees ($\gamma=2.96$),
triangulated outer-planar ($\gamma=4$),  and Apollonian
graphs ($\gamma=6.75$).
The recently updated upper bound \cite{Bonichon} comes
from counting unlabeled planar graphs.

The emergence of the logarithmic term in the entropy
expression Eq.(\ref{log}) is an integral part of our model.
Its coefficient $\beta=-\frac{5}{2}$ is not only
$\gamma$ independent, but most importantly, it is negative.
It automatically carries the vital minus sign, which allows us
to make contact with a variety of field theoretical calculations.
With the Cardy formula \cite {Cardy} serving as a light to guide
the way, first-order corrections to the Bekenstein-Hawking
entropy have been calculated \cite{Carlip,Majumdar,loopmodels}.
Despite very different physical assumptions, these
corrections seem to predominantly lead to $\beta=-\frac{3}{2}$.
Interestingly, the latter value would emerge had our graphs
been rooted trees (but they are not).

\section{VI. Exact solution (2+1 dimensions)}

By construction, our model has been exclusively designed
for a 3+1-dimensional spacetime, for which the black hole
horizon is 2 dimensional and has genus $0$.
Once an extra dimension is introduced, and the horizon
becomes a 3-dimensional surface ($S^3$ or $S^2\otimes S^1$),
tetrahedra replace the triangles, the planarity of the graphs
is gone, and their chordality, at least in the way defined, calls
for a nontrivial generalization. 

\begin{figure}[ht]
	\center
	\includegraphics[scale=0.4]{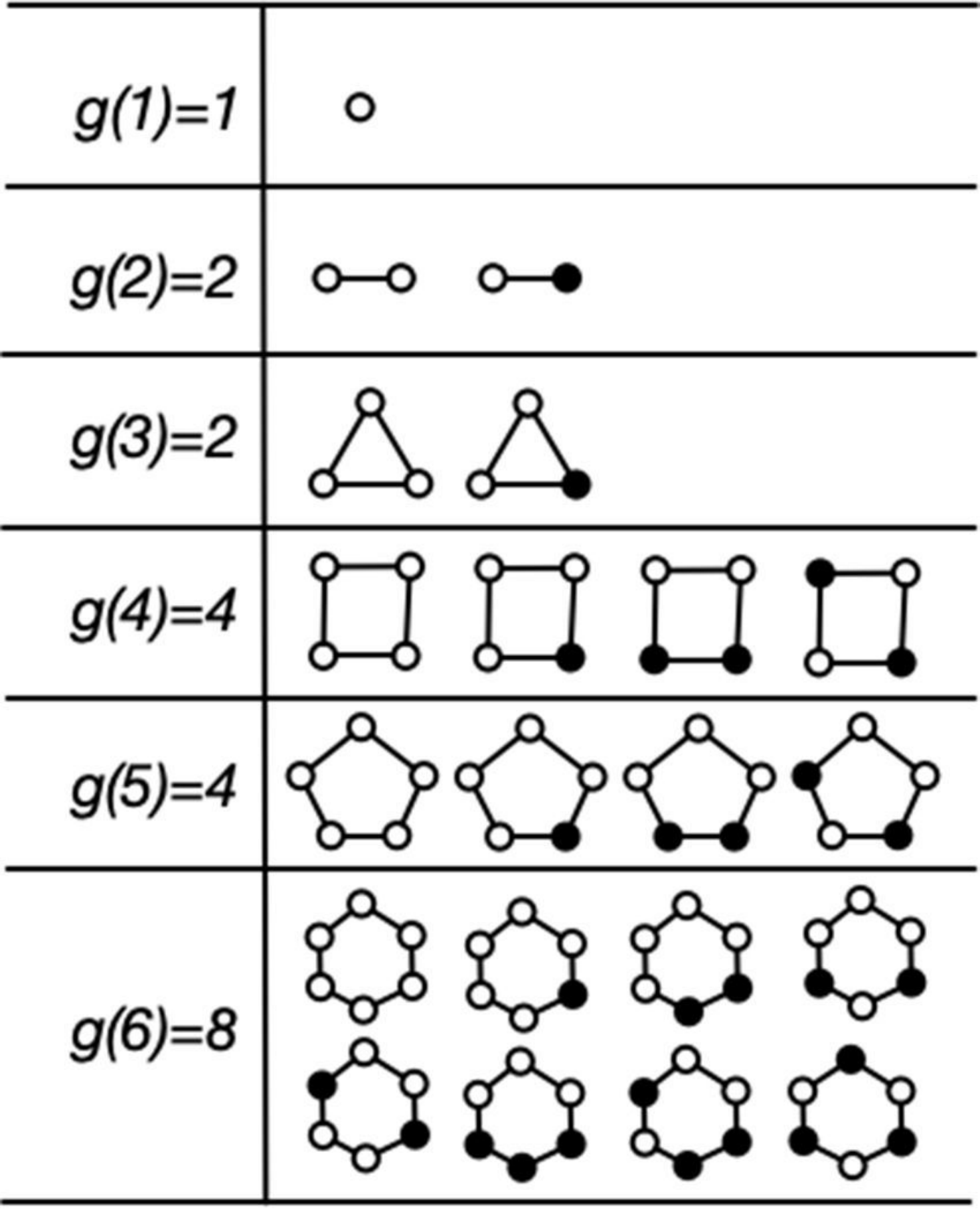}
	\caption{The integer sequence OEIS A000011:
	In 2+1 dimensions, graphs representing
	topologically distinct black hole microstates are
	free necklaces subject to a discrete
	$\circ \leftrightarrow \bullet$ symmetry, where
	$\circ=L$ bead and $\bullet=R$ bead.
	Each bead stands for a directed Planck unit arc.}
	\label{Necklace}
\end{figure}

On the other side, our model would naively and
wrongly suggest $g(n)=1$ for a 2+1-dimensional black
hole \cite{BTZ}, corresponding to tiling the now circular
horizon with $n$ equal-length unlabeled undirected
arcs.
The flexible shape unit areas previously introduced
have been replaced by firm unit arcs.
We are thus after a missing global ingredient,
characteristic of the $S^1$ topology, but that
does not have an $S^2$ analogue, and would similarly
allow for topologically distinct black hole microstates.
Indeed,  the topology of a circle naturally allows
for clockwise ($L$) and anticlockwise ($R$)
directions, a tenable feature that can be straightforwardly
translated into equal-length unlabeled
yet directed (arrow carrying) Planck unit arcs.
From the combinatorial point of view, we are then dealing
with a necklace of length $n=n_L+n_R$, composed of
two types of colored beads, $L$ beads $=\circ$ and
$R$ beads $=\bullet$ (beads of the same color are not
labeled differently), respectively.
Consistent with our topological approach, one
cannot locally tell $L$ from $R$ (chiralities, unlike
colors, do interchange once a necklace is flipped
over).
In other words, a discrete $L\leftrightarrow R$
symmetry applies (for example, $LLLR=RRRL$ 
should not be counted twice), as manifested in
Fig.\ref{Necklace}.
%% Fig Necklace %%%
%%%%%%

We count the number $g(n)$ of topologically distinct
necklaces by using Polya's generating function
method \cite{Polya}.
The main technical point is to prevent overcounting
of topologically equivalent configurations.
Hence, a central role in the calculation is played by the
discrete symmetries (its elements can be represented by
permutations) of the $n$ polygon.
Owing to these symmetries, the total number $g(n)$ of
necklaces, that is,
\begin{equation}
	g(n)=1,2,2,4,4,8,9,18,23,44,...,
\end{equation}
specified by the integer sequence OEIS A000011,
must be a function of all $\nu(n)$ divisors $d_i$ of $n$.While the basic formula, for $c$ colors($c=2$ in our case) is available \cite{necklace},
\begin{equation}
	N_n(c)=\frac{1}{2n}\sum_{i=1}^{\nu(n)}
	\phi({d_i})c^{\frac{n}{d_i}}+
	\left\{\begin{array}{lll}
	\frac{1}{2}c^{\frac{n+1}{2}} & n~ \text{odd}&
	\vspace{4pt}  \\
	 \frac{1}{4}(1+c) c^{\frac{n}{2}} & n~ \text{even}&  
	\end{array}\right. 
	\label{Polya}
\end{equation}
it has to be nontrivially adjusted to accommodate the
$L\leftrightarrow R$ symmetry imposed.
Eq.(\ref{Polya}) splits between $n$-odd and $n$-even terms
and introduces Euler's totient function $\phi(n)$
\cite{totient} (the number of integers $\leq n$
that are relatively primes to $n$).

The special case $n=$ odd prime, whose highlights
we now discuss in detail, is the simplest (no need to
calculate for each prime number individually) most
pedagogical case.
The associated point symmetry is the dihedral group
$D_{n}$.
It consists of $2n$ elements: $\phi(1)=1$ unity,
$\phi(n)=(n-1)$ rotations, and $n$ reflections.
These numbers $\{1,(n-1),n\}$, whose sum $2n$ matches
the order of the group, then enter as coefficients
into the cycle index of the group $D_n$, namely,
\begin{equation}
	Z[D_n]= \frac{1}{2n}\left(1 f_1^n+(n-1)f_n^1
	+n f_1^1 f_2^{\frac{n-1}{2}}\right)~.
\end{equation}
Following Polya, we now substitute
$f_p (L,R)=L^p+R^p$ to arrive at the correct
generating function $P_n (L,R)$.
The coefficient of the $L^p R^{n-p}$ term
($p=0,1,...,n$) in the polynomial expansion is
identified as the number of necklaces
consisting of $p$ L beads and $(n-p)$ R beads.
From here the way to $g(n)$ is already paved: to
be specific, $g(n)=\frac{1}{2}P_n(1,1)=\frac{1}{2}N_n(2)$,
with the factor $\frac{1}{2}$ reflecting the underlying
$L\leftrightarrow R$ symmetry.
Altogether, we derive an exact entropy formula
for a quantum black hole in 2+1 dimensions whose
circular horizon is tiled by an odd prime number $n$
of directed Planck unit arcs,
\begin{equation}
	S_{BH}(n)=k_B \log\left(
	\frac{1}{2n}\left(2^{n-1}
	+n 2^{\frac{n-1}{2}}+n-1\right)\right)
	\label{exact}
\end{equation}
The generalization, for an arbitrary integer $n$, reads
\begin{equation}
	S_{BH}(n)=k_B \log\left(
	\frac{1}{4n}\sum_{i=1}^{\nu(n)}
	\phi({2d_i})2^{\frac{n}{d_i}}+2^{[\frac{n-2}{2}]}\right)
\end{equation}
with $[x]$ denoting the floor function.
At the semiclassical (large $n$) limit, we once again
recover Eq.(\ref{SBH}), with the bonus being the
original Bekenstein-Mukhanov coefficient.
Typical of our model, it is automatically
accompanied by a logarithmic term,
characterized in this case by the $-1$ coefficient
\begin{equation}
	S_{BH}(n) \simeq k_B(n\log2-\log n-2\log2+...) ~.
	\label{nlog2}
\end{equation}
The asymptotic behavior holds for every integer
$n$, not just for primes, because the leading contribution
to $P_n(L,R)$ always comes from the $\frac{1}{2n}f_1^n$
term (associated with the largest divisor $n$).
%% Fig dS %%%
\begin{figure}[ht]
	\includegraphics[scale=0.5]{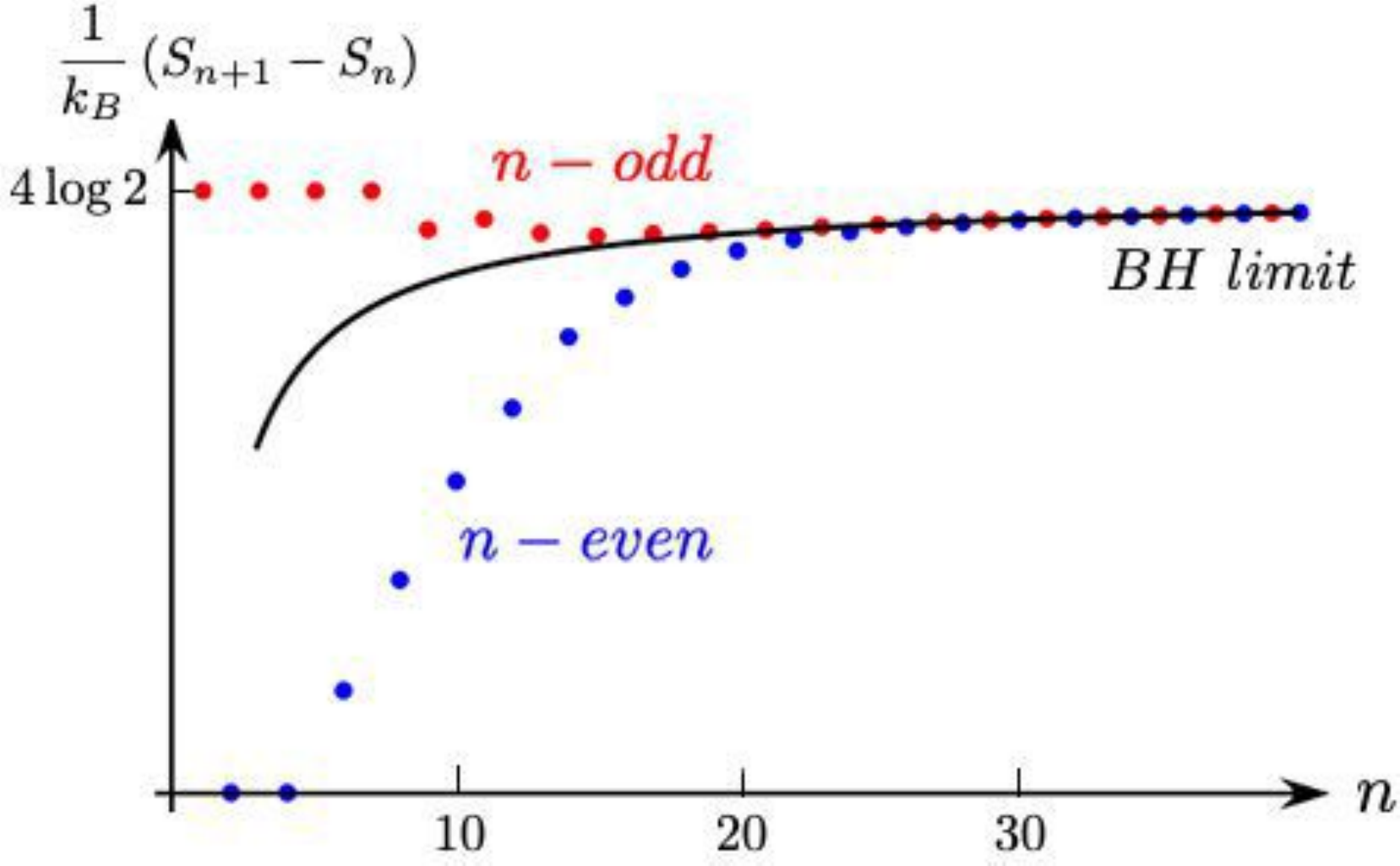}
	\caption{\label{dS} 
	The entropy increment $\Delta S$ per
	$\Delta n=1$ is plotted as a function of the
	number $n$ of Planck unit arcs which tile
	the circular horizon.
	Note the number theoretical Polya bifurcation
	into two branches, $n$-even (blue) and $n$-odd
	(red) respectively, sharing a common asymptotic
	limit (solid curve). }
\end{figure}
%%%%%

It is interesting to further study the deviation from the
Bekenstein-Hawking limit, in particular, for small $n$,
by plotting $\Delta S_n=S_{n+1}-S_n$, the amount of
entropy added by increasing the number of Planck arcs
by one unit.
The plot, see Fig. \ref{dS}, splits into two branches, even-$n$ and odd-$n$
branches, respectively.
As $n$ increases, the two branches merge to share a
common asymptotic behavior Eq.(\ref{nlog2}).

\section{VII. Conclusion}

Identifying and counting the elusive black hole microstates
is an open challenge in theoretical physics.
Counterintuitively, while invoking the familiar ingredient
of a fundamental Planck unit area, none of the individual units
play any local role in our model.
In fact, all Planck areas tiling the horizon are collectively
involved in what can be described as a global realization
of the "it from bit" phrase, with the topologically distinct
configurations resembling or even identified as the black
hole microstates.
This opens the door for graph theory and number
theory to enter black hole physics under
the auspices of the would be quantum gravity and/or
the universal complex network \cite{Bianchi}. 

If Eq.(\ref{n}) is not applicable to start with, our
model needs to be supplemented by field theoretic
ingredients or to be generalized graph theoretically. 
The first step would be to deal with Taub-NUT $S^3$
topology.
Regarding black hole phase transitions (topology change
or otherwise), our model cannot shed light on this aspect
at this stage.
Even the $n\rightarrow n-1$ quantum black hole
transition is not any clearer here than that given in the
standard spectral treatment of Bekenstein and Mukhanov. 
The fate of the "lost" Planck horizon unit area in the
process is under investigation and may hold the key to
future developments.

%%%%%%%%%%%%%%%%%% 
\section{Acknowledgments}
\acknowledgments
{I cordially thank Marc Noy (for helpful comments regarding
the critical exponent problem, and for substantiating the structure
of the asymptotic enumeration formula), Cyril Gavoille (for useful
comments on triangulations, outerplanar, and Apollonian graphs),
Falk H{\"u}ffner (for updating the OEIS A243787 integer sequence
using the TinyGraph software), and Shahar Hod (for a few critical
comments).
Special thanks to Judy Kupferman for her comments
and help in reorganizing the paper.}

\end{document}